# ON A NEW MULTICOMPUTER INTERCONNECTION TOPOLOGY FOR MASSIVELY PARALLEL SYSTEMS

C. R. Tripathy[1] and N. Adhikari[2]

Professor, Department of CSE, VSS University of Technology, Burla, Orissa, India
`write2na@gmail.com`
Asst. Professor, Department of CSE, PIET Rourkela, Orissa, India
`head.csepiet@gmail.com`

**ABSTRACT**

*This paper introduces a new interconnection network topology called Balanced Varietal Hypercube (BVH), suitable for massively parallel systems. The proposed topology being a hybrid structure retains almost all the attractive properties of Balanced Hypercube and Varietal Hypercube. The topology, various parameters, routing and broadcasting of Balanced Varietal Hypercube are presented. The performance of the Balanced Varietal Hypercube is compared with other networks. In terms of diameter, cost and average distance and reliability the proposed network is found to be better than the Hypercube, Balanced Hypercube and Varietal Hypercube. Also it is more reliable and cost-effective than Hypercube and Balanced Hypercube.*

## 1. INTRODUCTION

Parallel processing has assumed a crucial role in the field of supercomputing. It has overcome the various technological barriers and achieved high levels of performance. The most efficient way to achieve parallelism is to employ multicomputer system. The success of the multicomputer system completely relies on the underlying interconnection network which provides a communication medium among the various processors [9,18,29]. It also determines the overall performance of the system in terms of speed of execution and efficiency. The suitability of a network is judged in terms of cost, bandwidth, reliability, routing ,broadcasting, throughput and ease of implementation. Among the recent developments of various multicomputing networks, the Hypercube (HC) has enjoyed the highest popularity due to many of its attractive properties [11,19,31]. These properties include regularity, symmetry, small diameter, strong connectivity, recursive construction, partitionability and relatively small link complexity. In the literature variations of Hypercube topology has been proposed to further enhance some of its features. They include the Twisted cube [16] having less diameter than that of Hypercube, the Banyan Hypercube [2] and the Cube Connected Cycles [10]. In the Folded hypercube some complementary links are added .Thus it has still reduced diameter that is $\left\lceil \frac{n}{2} \right\rceil$ with degree (n+1) [3]. The Crossed cube is another improved variation of the Hypercube. It has smaller diameter $\left\lceil \frac{n+1}{2} \right\rceil$ than Hypercube with complex routing [15]. Another high performance –low cost architecture called the Incomplete crossed hypercube $CI_{n-m}^{n}$ is constructed by combining two crossed hypercube $CQ_n$ and $CQ_{n-m}$ for $1 \leq m \leq n$ [32]. It has shorter mean internode distance for large n. It is more useful than other incomplete networks.

The performance of Varietal hypercube has been compared with that of Hypercube, Folded hypercube, Twisted cube and Crossed cube in [6, 24]. The degree, average distance, and cost of Varietal cube is found to be the lowest among all these topologies. The Extended Hypercube is a hierarchical, expansive recursive structure with hypercube as the basic building blocks [17]. It has reduced diameter and average distance. With the use of Network controllers it has better routing properties than the hypercubes. Extended Varietal hypercube (EVH) is a recursive hierarchical structure and has still reduced diameter, average distance and constant degree of nodes[5,6]. Another variation of EVH is Extended varietal hypercube with crossed connections (EVHC) [27]. It overcomes the fault tolerant properties of EVH. Extended crossed cube is another similar type recursive, hierarchical network with well defined basic modules that is a crossed cube. It is having better features than the Extended varietal hypercube network.

One of the important class of Cayley graph, called the Star graph has been popular as an alternative to Hypercube [25,26]. It is a node symmetric and edge symmetric graph consisting of $n!$ number of nodes and $n!(n-1)/2$ number of edges. Some of the important features of Star graph are fault tolerance, partitionability, node disjoint paths and easy routing and broadcasting. Inspite of these attractive features, the Star network has a major disadvantage. It grows to its next higher dimension by a large value. Another alternative of Star called the Incomplete star has been introduced to eliminate this problem [28]. But the Incomplete star is a non symmetric and irregular graph. So it is not suitable to use in many practical systems. Recently Star-cube(n,m) network a variation of Star graph is introduced in [7]. The Star-cube also known as Cube-star is a hybrid network. The Star-cube is regular, vertex, edge-symmetric, maximally fault tolerant and cost effective. When compared with Star, the growth of Star-cube is comparatively small. The smallest possible structure contains 24 nodes with node degree 4.

Another variation of the Star graph called the Hierarchical star network, HS(n,n) is introduced as a two level interconnection network in [30]. The HS(n,n) network consists of *n!* number of modules where each module is a Star graph. So the HS network contains $(n!)^2$ nodes with node degree n. The modules are interconnected with additional edges. The size of the network grows at a very high rate. When n is 3 the network size is 36 but when n is 4, the network contains 576 nodes. This significant gap in the two consecutive sizes of Hierarchical Star becomes a major disadvantage. Another disadvantage of Hierarchical star is that the dimension cannot take any values of n like Starcube. It only takes values like (3,3), (4,4), (5,5) etc.

Irrespective of the network type with increasing number of processors, the system reliability is also expected to decrease. For this reason alternate fault tolerant features are to be introduced in the network. The fault tolerance aspect of the Balanced hypercube (BH) is proved to be better than that of the Hypercube [12,14,20,21]. Each processor in $BH_n$ has a backup processor that is having the same set of neighbouring nodes. The Balanced hypercube is beneficial for parallel processing in terms of reduced diameter only when the dimension is odd.

However the performance parameters such as reliability, fault tolerance, cost effectiveness and the time-cost effectiveness are some of the important aspects that need to be addressed while designing any large scale parallel system [1,4,5,6,8].

For this reason there has always been raising demands for design of a versatile interconnection network with efficient communication, better reliability, improved fault tolerance and reduced cost. The present paper attempts to meet the above demands and proposes a new network topology called the Balanced Varietal Hypercube (BVH). The proposed topology is a hybrid structure of the Balanced Hypercube ($BH_n$) and the Varietal Hypercube ($VQ_n$). The BVH is built on the basic structure of the Varietal hypercube and BH. It inherits the merits of fault-tolerance from BH. In addition, the BVH has got a reduced diameter, optimal average distance with less cost. It is also a load balanced graph.

The current paper is organized as follows. Section 2 presents the architectural details. Topological properties of the proposed Balanced Varietal hypercube are presented and discussed in Section 3. The routing and broadcasting aspects are discussed in Section 4. Performance comparison is carried out in Section 5. The Section 6 concludes the paper.

## 2. ARCHITECTURAL DETAILS

This section describes the topological features of the Varietal Hypercube and the Balanced Hypercube The above said interconnection network topologies are described using graph theoretical terminologies and notations.

### 2.1 Varietal Hypercube

The Varietal Hypercube is a variation of Hypercube with reduced diameter and average distance [24]. An n-dimensional Varietal Hypercube ($VQ_n$) is constructed from two numbers of (n-1) dimensional Varietal Hypercubes in a way similar to that of the Hypercube with some modifications in connections. The connections are as follows: $VQ_1$ is a complete graph of two vertices with address 0 and 1. For n>1, $VQ_n$ is constructed from $VQ_{n-1}^0$ and $VQ_{n-1}^1$ according to the rule: a vertex u with node address $(0, u_{n-1} u_{n-2} u_{n-3} \ldots u_1)$ from $VQ_{n-1}^0$ and a vertex v with node address $(1, v_{n-1} v_{n-2} v_{n-3} \ldots v_1)$ from $VQ_{n-1}^1$ are adjacent in $VQ_n$ if and only if

1) $u_{n-1} u_{n-2} u_{n-3} \ldots u_1 = v_{n-1} v_{n-2} v_{n-3} \ldots v_1$, if n=3k or

2) $u_{n-3} \ldots u_1 = v_{n-3} \ldots v_1$ and $(u_{n-1} u_{n-2}, v_{n-1} v_{n-2}) \in \{(00,00),(01,01),(10,11),(11,10)\}$, if n=3k. The Varietal Hypercube of dimension 3 is shown in Fig. 1.

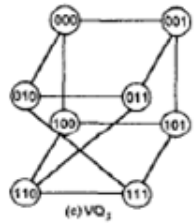

Figure 1: Varietal Hypercube of dimension 1,2 and 3

### 2.2. Balanced Hypercube

The Balanced Hypercube network of dimension n ($BH_n$) is a load balanced graph having $2^{2n}$ nodes [14,21]. Each vertex of $BH_n$ has a unique n-component vector on {0, 1, 2, and 3} for its label such as $(a_0, a_1, \ldots, a_{n-1})$. A vertex u having label $(a_0 a_1 \ldots a_{n-1})$ is adjacent to the following 2n vertices for $1 \leq i \leq n-1$,

$((a_0 + 1) \bmod 4, a_1, \ldots, a_{i-1}, a_i, a_{i+1}, \ldots a_{n-1})$, $((a_0 - 1) \bmod 4, a_1, \ldots, a_{i-1}, a_i, a_{i+1}, \ldots a_{n-1})$, $((a_0 + 1) \bmod 4, a_1, \ldots, a_{i-1}, (a_i + (-1)^{a_0}) \bmod 4, a_{i+1}, \ldots a_{n-1})$ and $((a_0 - 1) \bmod 4, a_1, \ldots, a_{i-1}, (a_i + (-1)^{a_0}) \bmod 4, a_{i+1}, \ldots a_{n-1})$. The Balanced Hypercubes of dimension 3 is shown in Fig.2. The $BH_n$ can be constructed from four copies of $BH_{n-1}$ by adding a new edge in the nth dimension of every vertex in $BH_n$.

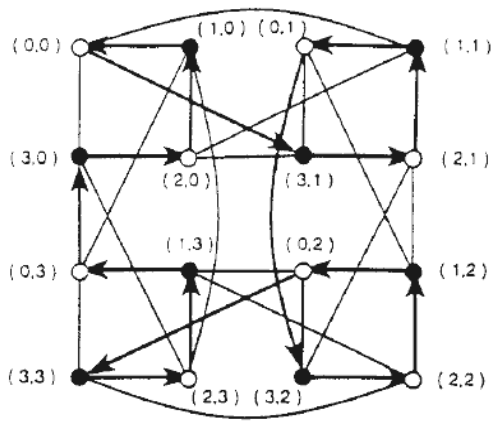

Figure 2: Balanced Hypercube of dimension 3

## 3. PROPOSED TOPOLOGY

The present section is devoted towards providing the topological details of the proposed topology.

### 3.1 Balanced Varietal Hypercube

Let $G=\{V,E\}$ be a finite, undirected graph with set of nodes $V$ and set of edges $E$. A node in $V$ represents a processor and an edge in $E$ represents a communication link between two processors. If an edge $e=(u,v) \in E$, then the nodes $u$ and $v$ are adjacent. For each node $v$ there exists another node $v'$ such that $v$ and $v'$ have same adjacent nodes. The pair $v$ and $v'$ are called matching pair. A task can be scheduled to both $v$ and $v'$ in such a way that one copy is active and the other one is passive. If node $v$ fails, its task can simply be shifted to node $v'$ by activating copies of these tasks in $v'$. All the other tasks running on other nodes need not be reassigned to keep the adjacency property, that is two tasks those are adjacent are still adjacent after the reconfiguration. It is possible to have an active task running on node $v$ with its backup in $v'$, while having another active task on $v'$ and its backup on node $v$. The degree $d(v)$ of node $v$ is equal to the number of edges in G which are incident on $v$. The diameter of G is the maximum distance between two nodes in G over all pairs of nodes. The Balanced varietal hypercube of different dimensions are shown in Fig. 3 and 4.

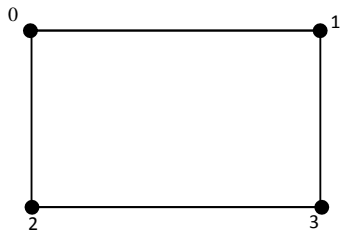

Figure 3: Balanced Varietal Hypercube of dimension 1

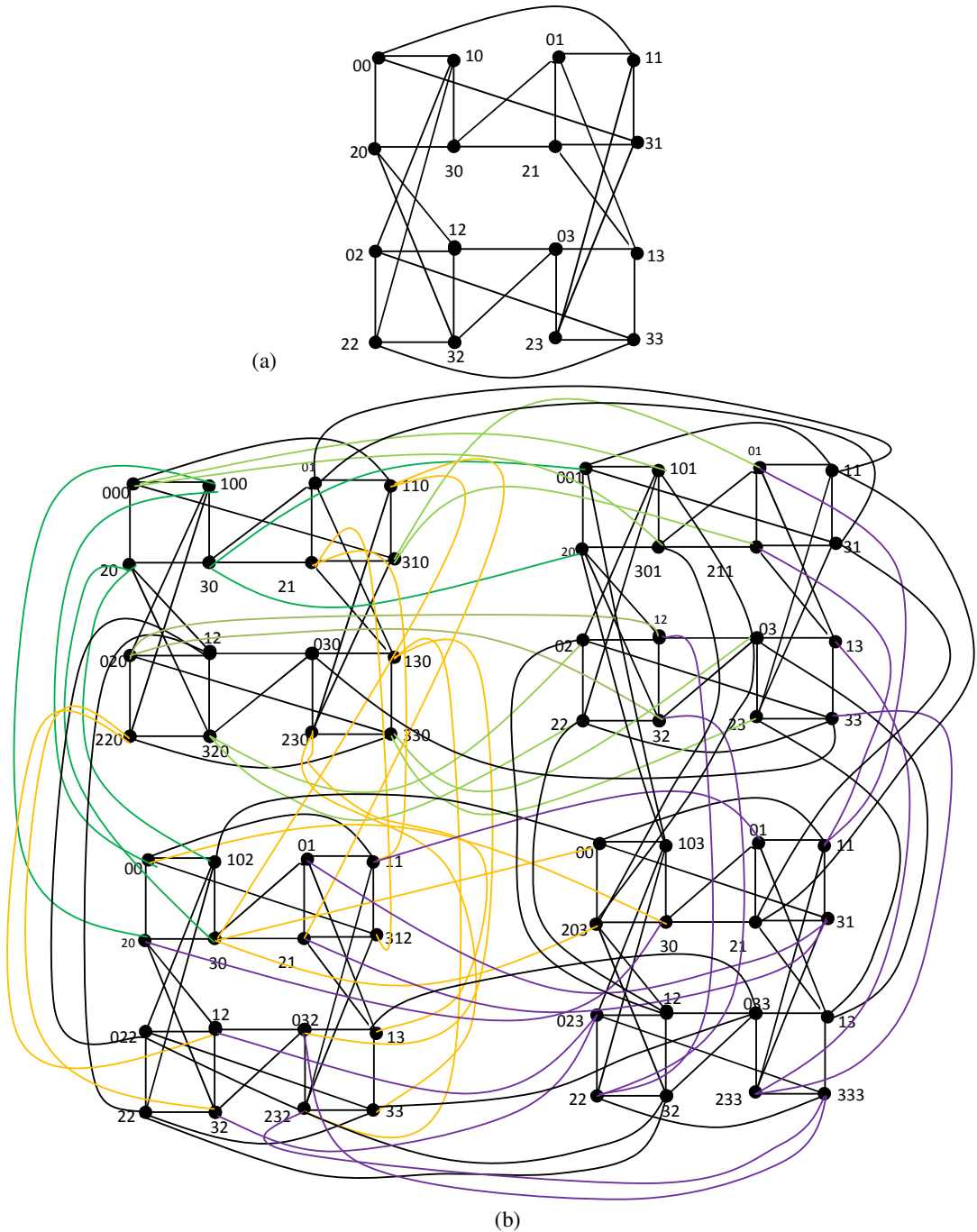

Figure 4: Balanced Varietal Hypercube of dimension 2 and 3,(a)BVH$_2$ (b) BVH$_3$

*Definition 3.1*

An n-dimensional Balanced Varietal Hypercube (BVH$_n$) consists of $2^{2n}$ nodes each of which is represented by the address $(a_0,a_1,a_2,a_3,.....a_i,...a_{n-1})$ where $a_i \in \{0,1,2,3\}$ and $0 \leq i \leq n-1$. Every node $(a_0,a_1,a_2,a_3,.....a_i,...a_{n-1})$ connects the following 2n nodes, which are divided into two categories: a) inner nodes and b) outer nodes. In an n dimensional Balanced Varietal hypercube BVH$_n$ each unit is connected to others through hyperlinks.

   a) Inner node:
Case I: When $a_0$ is even,
   (i) $<(a_0+1) \bmod 4, a_1, a_2..... a_{n-1}>$
   (ii) $<(a_0-2) \bmod 4, a_1, a_2..... a_{n-1}>$
Case II: When $a_0$ is odd,
   (i) $<(a_0-1) \bmod 4, a_1, a_2..... a_{n-1}>$
   (ii) $<(a_0+2) \bmod 4, a_1, a_2..... a_{n-1}>$
   b) Outer node:
Case I: When $a_0=0,3$;
   (i) For '$a_i$' = 0
   $<(a_0+1)_{\bmod 4}, a_1,....,(a_i+1)_{\bmod 4} a_2,....,a_{n-1}>$
   $<(a_0-1)_{\bmod 4}, a_1,....,(a_i+1)_{\bmod 4} a_2,....,a_{n-1}>$
   (ii)   For '$a_i$' = 3
   $<(a_0+1)_{\bmod 4}, a_1,....,(a_i-1)_{\bmod 4},....,a_{n-1}>$
   $<(a_0-1)_{\bmod 4}, a_1,....,(a_i-1)_{\bmod 4},....,a_{n-1}>$
Case II: when $a_0=1,2$ and $a_i= 0,3$
   $<(a_0+1)_{\bmod 4}, a_1,....,(a_i+2)_{\bmod 4},....,a_{n-1}>$
   $<(a_0-1)_{\bmod 4}, a_1,....,(a_i+2)_{\bmod 4},....,a_{n-1}>$
Case III: when $a_0=0,1$
   (i) For $a_i=1$
   $<(a_0+1)_{\bmod 4}, a_1,....,(a_i+2)_{\bmod 4},....,a_{n-1}>$
   $<(a_0-1)_{\bmod 4}, a_1,....,(a_i-1)_{\bmod 4},....,a_{n-1}>$
   (ii) For $a_i=2$
   $<(a_0+1)_{\bmod 4}, a_1,....,(a_i+2)_{\bmod 4},....,a_{n-1}>$
   $<(a_0+1)_{\bmod 4}, a_1,....,(a_i+2)_{\bmod 4},....,a_{n-1}>$
Case IV: when $a_0=2,3$
   (i) For $a_i=1$
   $<(a_0+1)_{\bmod 4}, a_1,....,(a_i-1)_{\bmod 4},....,a_{n-1}>$
   $<(a_0-1)_{\bmod 4}, a_1,....,(a_i+2)_{\bmod 4},....,a_{n-1}>$
   (ii) For $a_i=2$
   $<(a_0+1)_{\bmod 4}, a_1,....,(a_i+1)_{\bmod 4},....,a_{n-1}>$
   $<(a_0-1)_{\bmod 4}, a_1,....,(a_i+2)_{\bmod 4},....,a_{n-1}>$

## 3.2 Degree

The degree of a node in a graph is defined as the total number of edges connected to that node. Similarly the degree of a network is defined as the largest degree of all the vertices in its graph representation.

***Theorem3.1:*** The degree of any node in the Balanced varietal hypercube of dimension n is equal to 2n.

***Proof:*** From the Definition 3.1 it is clear that $BVH_1$ is constructed from four nodes and the number of edges connected to each node is 2. A balanced varietal hypercube of any dimension $BVH_n$ is constructed from four $BVH_{n-1}$ with each node having two extra connections as shown in Fig.4. So when the dimension is increased by one, the number of extra connections made to each node is increased by 2. Hence, the theorem is proved.

### 3.3 Number of Nodes

In a finite undirected graph $G=(V,E)$, $V$ represents the node set and $E$ represents the edge set. Normally a node in $V$ represents a processor and an edge in $E$ corresponds to a communication link connecting two processors.

***Theorem3.2:*** An n-dimensional Balanced varietal hypercube has $2^{2n}$ nodes.

***Proof:*** A Balanced varietal hypercube is a load balanced graph, that is for every node there exist another node such that these two nodes are having same adjacent nodes. Hence an n-dimensional Balanced varietal hypercube is very much similar to a varietal hypercube of dimension 2n, and the number of nodes is same as that of n-cube [23].

Lemma1: A graph $G=(V,E)$ is an n-cube if and only if

  a) V has $2^n$ vertices.
  b) Every vertex has degree n.
  c) G is connected.
  d) Any two adjacent nodes A and B are such that the nodes adjacent to A and those adjacent to B are linked in a one-to-one fashion.

For a one dimensional Balanced varietal hypercube, shown in Fig. 3, the number of nodes is equal to $2^{2*1}=4$ nodes. For a two dimensional Balanced varietal hypercube shown in Fig.4, the total number of nodes are equal to $2^{2*2}=16$. Similarly, for a three dimensional balanced varietal hypercube the total number of nodes is equal to $2^{2*3}=64$.

Hence, by induction it can be proved that the n-dimensional $BVH_n$ has $2^{2n}$ nodes.

### 3.4 Number of Edges

An edge represents a communication link between two processors in a network. If an edge $e=(u,v) \in E$, then the nodes u and v are adjacent.

***Theorem3.3:*** An n-dimensional Balanced varietal hypercube has $n*2^{2n}$ edges.

***Proof:*** From Theorem 3.2, an n-dimensional Balanced varietal hypercube has $2^{2n}$ nodes. According to Theorem 3.1 the degree of any node in an n-dimensional BVH is 2n. But a link is shared by two nodes as shown in Fig. 3. Therefore the total number of links or edges for $BVH_n$ is $2n*2^{2n}/2 = n2^{2n}$

### 3.5 Diameter

The diameter is considered to be the most important parameter of any network. The distance d(u,v) between two distinct vertices is the length of the shortest path between these vertices. The diameter of G, denoted as D(G) is defined to be the maximum of these distances. Since the diameter is the worst case distance in a graph, it reflects how long it would take for a node to broadcast message to all other nodes.

*Theorem 3.4 :* The diameter of an n-dimensional Balanced varietal hypercube is

    i.      2n for n=1
    ii.     $\left\lceil n + \frac{n}{2} \right\rceil$ for n> 1.

*Proof:* The theorem is proved by induction.

For n=1, using Fig 3. It is clear that the diameter of $BVH_1$ is 2. For n=2, as shown in Fig. 4(a), the maximum of the shortest distance between two nodes is $\left\lceil 2 + \frac{2}{2} \right\rceil = 3$. The distance of each node is calculated from every other node. For $BVH_3$, the distance is $\left\lceil 3 + \frac{3}{2} \right\rceil = 4$.

Let $u=(a_0,a_1,a_2,....a_{n-1})$ and $v=(b_0,b_1,b_2,...b_{n-1})$ be two nodes in an n-dimensional balanced varietal hypercube. When $a_{n-1} \neq b_{n-1}$ it can be considered that u and v are on two adjacent BVHs of dimension n-1. Hence, the distance between them is $\left\lceil n + \frac{n}{2} \right\rceil$ as n>1. Hence the result is true for n.

When $a_{n-1} = b_{n-1}$, then the nodes are on the same $BVH_{n-1}$. Hence the distance between them is less than $\left\lceil n + \frac{n}{2} \right\rceil$ and equal to $\left\lceil \frac{2n-3}{2} \right\rceil$.

### 3.6 Average Distance

In a loosely coupled distributed system, while executing any parallel algorithm message traffic between processors takes on a distribution fairly close to uniform distribution. The average distance conveys the actual performance of the network better in practice. The summation of distance of all nodes from a given node over the total number of nodes determines the average distance of the network [19,31].

*Theorem 3.5 :* In the Balanced varietal hypercube the average distance $\bar{d}(BVH_n)$ is given by

$\bar{d}(BVH_n) = \frac{1}{2^{2n}} \sum d[(0,0,0),k]$ ; all node in $BVH_n$

*Proof:* The total number of nodes in $BVH_n$ is $2^{2n}$. The average distance is the ratio of sum of distances of all nodes from a given node to the total number of nodes.

### 3.7 Message Traffic Density

The performance of a network in handling the message traffic can be analysed by assuming that each node is sending a message to a node at distance $\bar{d}$ on the average. An efficient network should have a wide enough bandwidth to handle the resulting traffic so that the message traffic density is the minimum.

*Theorem 3.6:* The message traffic density for an n-dimensional Balanced varietal hypercube is
$\frac{\bar{d}(BVH_n) 2^{2n}}{n 2^{2n}}$

*Proof:* As discussed earlier, the message traffic density can be calculated if we know the average distance, the total number of nodes and the total number of edges. From Theorem 3.2, the number of nodes in a BVH of dimension n is $2^{2n}$. From Theorem 3.3, the number of edges in $BHV_n$ is $n*2^{2n}$. Using the average distance of a n-dimensional BVH can be calculated. Hence

Message traffic density $= \dfrac{Avg.Distance * No.of\ nodes}{No.of\ links}$

$= \dfrac{d(BVH_n) 2^{2n}}{n 2^{2n}}$

## 3.8 Cost

Cost is an important factor as far as an interconnection network is concerned. The topology which possesses minimum cost is treated as the best candidate. Cost factor of a network is the product of degree and diameter.

*Theorem 3.7:* The cost of an n-dimensional balanced varietal hypercube is given by $2n*\left\lceil n + \dfrac{n}{2} \right\rceil$.

*Proof:* The degree of an n-dimensional BVH is 2n. The diameter is $\left\lceil n + \dfrac{n}{2} \right\rceil$. Since the cost is product of degree and diameter, hence for a $BVH_n$

Cost= degree* diameter= $2n*\left\lceil n + \dfrac{n}{2} \right\rceil$ for n>1.

## 3.9 Node-disjoint Path

The Node-disjoint path defines in how many ways two nodes can be linked without any common node. The Node-disjoint paths are to be considered quite important while designing an interconnection network.

*Theorem 3.8:* For any pair of nodes in an n-dimensional Balanced varietal hypercube, there exists 2n disjoint paths between them.

*Proof:* For one dimensional BVH, the Node-disjoint paths between any two nodes are equal to 2*1=2. From Fig. 3, considering nodes 0 and 3

    Path 1: 0-1-3
    Path 2: 0-2-3

In two dimensional BVH, the Node-disjoint paths will be 2*2=4. For example, from node (0,0) and (3,3) the different paths are

    Path1: 0,0-1,1-2,3-3,3
    Path 2: 0,0-1,0-2,2-3,3
    Path3 :0,0-3,1-2,1-3,3
    Path 4:0,0-2,0-1,2-0,2-3,3

Similarly from Fig.4, the Node-disjoint paths between (0,0,0) and (3,3,0) are

Path 1: (0,0,0)-- (1,0,0)-- (0,2,0)-- (3,3,0)
Path 2: (0,0,0)-- (3,1,0)-- (2,3,0)-- (3,3,0)
Path3:(0,0,0)--(2,0,0)--(1,2,0)--(3,2,0)--(2,2,0)-- (3,3,0)
Path 4: (0,0,0)--(1,1,0)--(0,1,0)--(2,1,0)-- (1,3,0)-- (3,3,0)

Path5: (0,0,0)--(1,0,1)-- (0,2,1)-- (1,2,1)-- (0,3,1)
Path 6: (0,0,0)--(3,0,1)--(0,1,1)-- (2,3,1)-- (3,3,0)

So there are 2*3=6 different paths for a 3-dimensional BVH. By induction it can be proved that for an n-dimensional BVH there will be 2n paths.

## 4. MESSAGE ROUTING

In multicomputer networks, communication is an important issue regarding how the processor can exchange message efficiently and reliably. An optimal routing algorithm aims to find the shortest path between two nodes communicating with each other.

### 4.1 Routing

In routing process, each processor along the path considers itself as the source and forwards the message to a neighbouring node one step closer to the destination. The algorithm consists of a left to right scan of source and destination address. Let $r$ be the right most differing bit (quarternary) position. The numbers to the right of $u_r$ is not to be considered as they lie on the same $BVH_r$. Since the diameter of $BVH_1$ is 2 there is atleast one vertex which is a common neighbour of $u_r$ and $v_r$. If $d$ is an element such that $d$ neighbour of $u_r$ is also a neighbour of $v_r$. Then $d$ is choosen such that $u_r=v_r$. Then in the next step $d_1$ is choosen such that $u_{r-1}=v_{r-1}$. This process continues until $u=v$.

*Algorithm:Procedure Route(u,v)*
*{r: right most differing bit position*
*d:choice such that $du_r=v_r$*
*route to d-neighbour else*
*route to r-neighbour (k and v are adjacent)*
*if (u and v are not adjacent) then*
*$d_1$=choice that $du_{r-1}=dv_{r-1}$*
*route to $d_1$ neighbour*
*}*

this process continues till $u_0,u_1,u_2,...u_{r-1},u_r=v_0,v_1,v_2,...v_{r-1},v_r$.

Finally, u=v that is source = destination.

### 4.2 Broadcasting

Broadcasting is the process of information dissemination in a communication network by which a message originated at a node is transmitted to all other nodes in the network. The broadcast primitive finds wide application in the control of distributed systems and in parallel computing. For instance, in computer networks, there are many tasks, such as scheduling and updating other processors in order to continue the processing.

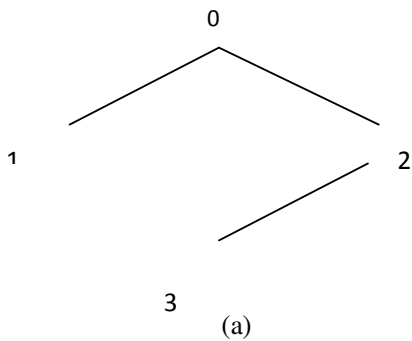

(a)

An optimal one-to-all broadcast algorithm is presented for $BVH_n$ assuming that concurrent communication through all ports of each processor is possible. It consists of (n+1) steps.

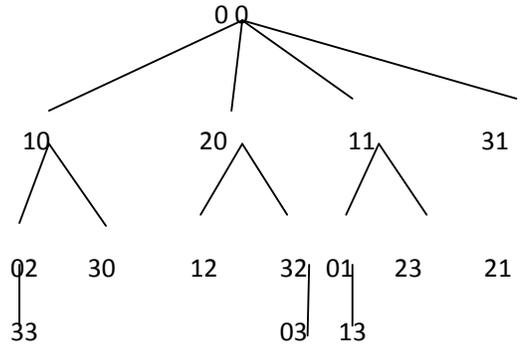

Figure 5 : Broadcasting (a) in $BVH_1$ (b): in $BVH_2$

*Lemma 1:*

The oriented versions of the trees $ST_i$ obtained by directing arcs from parent to child for i=0,1,... d-1 are pair wise arc disjoint.

Procedure *Broadcast(u,n):*

*Step1: send message to 2n neighbours of u*
*Step 2: one of 2n nodes sends message to its 2n-1 neighbours. Then n nodes from the rest nodes send message to their (2n-2) neighbours.*
*Step 3: continue step 2 till all the nodes get the message.*
*Step 4: end*

This has been illustrated in Fig.5 (a) and (b) for one dimensional and two dimensional BHV respectively.

## 5. PERFORMANCE ANALYSIS

All interconnection topologies may not be suitable for each task. Therefore, before selecting a particular topology, it is important to compare its performance with its predecessors. The present section is a systematic attempt to compare the various performance parameters of the proposed BVH with that of VH, BH and HC. The various performance parameters analysed below are: degree, diameter, cost, average distance, cost effectiveness, time cost effectiveness and reliability.

### 5.1 Comparison of Topological Parameters

In this subsection, the various topological parameters of the BVH is compared with Hypercube, Varietal hypercube and Balanced hypercube.

The Fig. 6 provides a comparative illustration of the diameter of the BVH. The diameter of BVH is observed to lie between that of the Varietal hypercube and the Balanced hypercube. In case of BVH, the diameter is slightly more than that of the Varietal hypercube, however, it is very less than that of Balanced hypercube there by reducing worst case delay in communication.

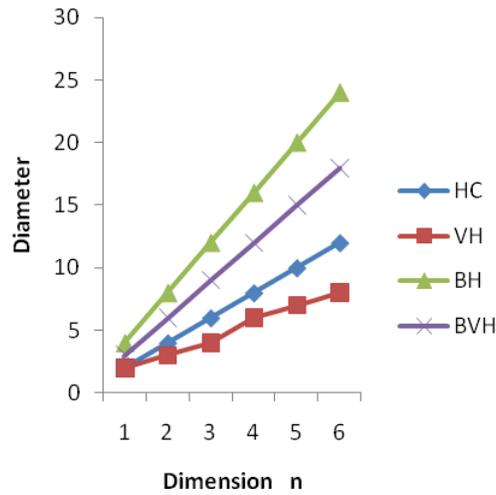

Figure 6: Comparison of Diameter

Since the BVH provides a lower diameter with the degree remaining the same as compared to BH and HC, the cost factor of the BVH is much less than that of BH and hypercube. The Fig. 7 compares the cost versus the dimension for hypercube, varietal hypercube, balanced hypercube and balanced varietal hypercube.

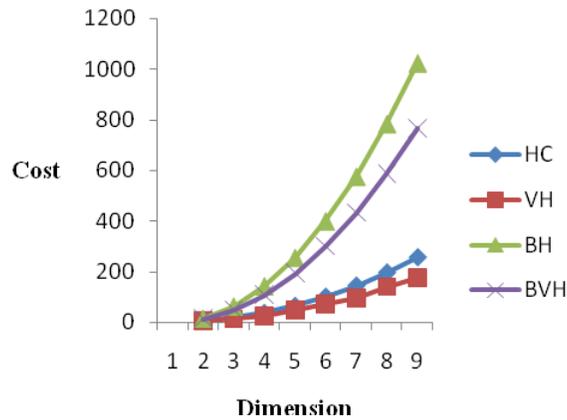

Figure 7: Comparison of cost of BVH

The average distance of a network reflects the actual performance of a network in a better way. The Table 1 and Fig. 8 show the superiority of BVH over its counterpart BH in terms of the average distance.

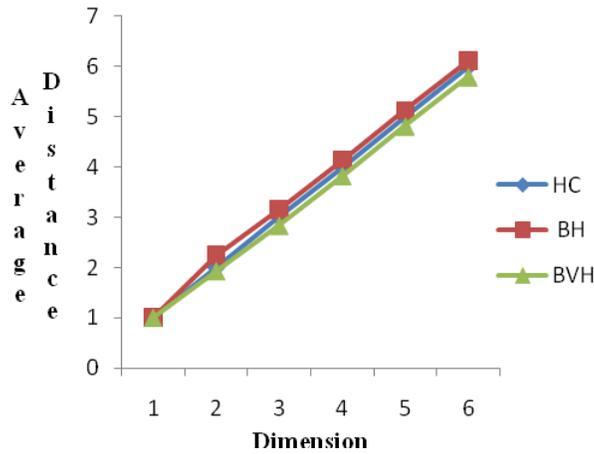

Figure 8 : Comparison of Average Distance of BVH

Table 1 : Comparison of Average Distance of BVH with other Networks

| Size n | $\bar{d}\,(HC_n)$ | $\bar{d}\,(BH_n)$ | $\bar{d}\,(BVH_n)$ |
|---|---|---|---|
| 1 | 1 | 1 | 1 |
| 2 | 1 | 2.25 | 1.93 |
| 3 | 1.5 | 3.156 | 2.83 |
| 4 | 2 | 4.14 | 3.82 |
| 5 | 2.5 | 5.12 | 4.81 |
| 6 | 3 | 6.11 | 5.79 |

## 5.2 . Cost Effectiveness Factor

The total cost of a multicomputer system comprises of the cost of the processors as well as the cost of the communication links. Usually, the number of links is a function of the number of processors. Thus, the earlier methods of performance evaluation by speedup and efficiency are inadequate. The cost effectiveness factor gives more insight to the performance of parallel systems that uses parallel algorithms [8].

Cost effectiveness of the BVH is a product of two terms, one characterises the architecture and the other corresponds to the efficiency of the algorithm. Therefore, the Cost effectiveness factor, CEF(p) for the proposed system is the ratio of cost effectiveness CE(p) to the efficiency E(p) where p is the total number of processors in the system. Here the number of links is a function of the number of nodes in the system.

The CEF of BVH is given by

$$\text{CEF}(p) = \frac{CE(p)}{E(p)} = \frac{1}{1+\rho g(p)} \quad (1)$$

where

$$\rho = \frac{c_l}{c_p} = \frac{\text{cost of a link}}{\text{cost of a processor}}$$

and

$$g(p) = f(p)/p$$

$f(p)$ gives the number of links as a function of $p$, the total number of nodes and n, the diameter of the network.

For $BVH_n$, $p = 2^{2n}$. The total number of links is given by
$$E = n2^{2n} = f(p).$$
$$\text{Hence }, g(p) = \frac{f(p)}{p} = \frac{n2^{2n}}{2^{2n}} = n \qquad (2)$$

Now using Eq. (2) in (1), we can have
$$\text{CEF}(p) = \frac{1}{1+\rho n} \qquad (3)$$

CEF enables the comparison of different parallel algorithms in different multicomputer architectures to determine the most cost effective combination of algorithm and architecture.

Table 2 : Cost Effectiveness Factor for BVH

| Dimension | Nodes | $\rho = 0.1$ | $\rho = 0.2$ | $\rho = 0.3$ |
|---|---|---|---|---|
| 1 | 4 | 0.909 | 0.833 | 0.769 |
| 2 | 16 | 0.833 | 0.714 | 0.625 |
| 3 | 64 | 0.769 | 0.625 | 0.526 |
| 4 | 256 | 0.714 | 0.555 | 0.454 |
| 5 | 1024 | 0.666 | 0.500 | 0.400 |
| 6 | 4096 | 0.625 | 0.454 | 0.357 |

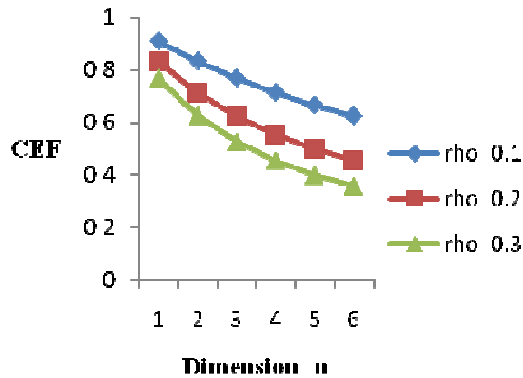

Figure 9: Comparison of Cost Effectiveness Factor of BVH

The Table 2 presents the computed values of CEF for the BVH. Figure 9 shows variations of CEF with the dimension of the proposed parallel system. It is a monotonically decreasing function of $p$ like the hypercube [8]. Thus, when the network size grows, it becomes less and less cost effective.

## 5.3 Time Cost Effectiveness Factor

The consideration of time factor is essential in evaluation of performance of a parallel system. The Time cost effectiveness factor (TCEF) takes into account the time factor in addition to the cost effectiveness factor considered in the above paragraph. It considers the situation where a faster solution to a problem is more rewarding than a slower solution [8].

$$\text{TCEF}(p, T_p) = \frac{1 + \sigma T_1^{\alpha-1}}{1 + \rho g(p) + \frac{T_1^{\alpha-1} \sigma}{p}} \qquad (4)$$

where $T_1$ is the time required to solve the problem by a single processor using the fastest sequential algorithm, $T_p$ is the time required to solve the problem by a parallel algorithm using a multicomputer system having $p$ processors and $\sigma$ is the ratio of cost of penalty with the cost of processors. For linear time penalty in $T_p$, $\alpha$ is choosen as 1.

Now using Eq. (2) in Eq.(4) the TCEF for $BVH_n$ is given by

$$\text{TCEF}(p, T_p) = \frac{1+1}{1+\sigma n + \frac{1}{2^{2n}}} \qquad (5)$$

Table 3 : TCEF for The BVH network

| Dimension | Nodes | $\rho = 0.1$ | $\rho = 0.2$ | $\rho = 0.3$ |
|---|---|---|---|---|
| 1 | 4 | 1.48148 | 1.37931 | 1.29032 |
| 2 | 16 | 1.58415 | 1.36752 | 1.20300 |
| 3 | 64 | 1.52019 | 1.23791 | 1.04404 |
| 4 | 256 | 1.42459 | 1.1087 | 0.90748 |
| 5 | 1024 | 1.33246 | 0.9995 | 0.79968 |
| 6 | 4096 | 1.249809 | 0.90899 | 0.71422 |

The computed values of TCEF for $BVH_n$ is shown in Table 3 keeping the value of $\rho$ costant and $\sigma$ value varied. The TCEF for the networks of varying sizes is shown in Fig. 10.

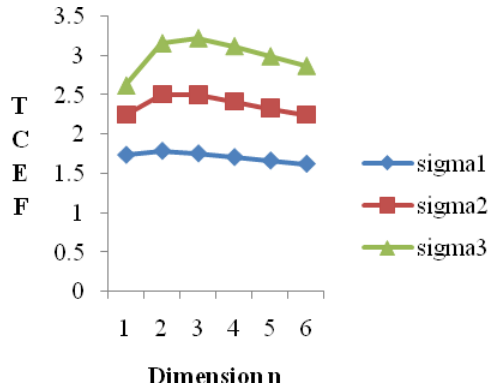

Figure 10: Comparison of TCEF

From the figure it is clear that the network is most suitable when the number of processor lies between 16 to 64.

## 5.4 Reliabilty Analysis

The assessment of reliability is very important for critical systems like the parallel systems. Reliability is the conditional probability that a system will survive in an interval $(0, \Delta t)$, given that it was operational at time t=0. The reliability of an electronic component ($R_t$) of the system is given by

$$R_t = e^{-\lambda t} \qquad (6)$$

where $\lambda$ is the failure rate of the component and t is the mission time.

Reliability of a network is dependent on the reliability of its components at the hardware level. It decreases in an exponential manner with time. Hence, the reliability is not only dependent on the topology but also on time. From the topological point of view, reliability issues have been addressed by different researchers [4,5,22]. For simplicity, two terminal reliability or simply terminal reliability is considered here.

The terminal reliability is defined as the reliability between any two specified nodes termed as source and destination. The total number of node disjoint paths, as well as number of links and nodes involved in a particular path are important for evaluation of reliability. The reliability analysis has been carried out following a method called sum of disjoint products (SDP) [21,22]. Using the said method the probability of each term is found out separately which is then added together to get the exact two-terminal reliability. For calculating the terminal reliability (TR) between two given nodes of a network, the reliability of each node as well as edge are also considered.

Terminal reliability between a pair of nodes is given by

$$TR = 1 - [(1 - R_l^m R_p^n)^k (1 - R_l^{m'} R_p^{n'})^{k'}] \qquad (7)$$

where, $R_l$ =Reliability of each link

$R_p$=Reliability of each processor (node) where there are $k'$ paths with $m'$ number of links and $n'$ number of processors in each path.

### 5.4.1. Reliability of BVH$_2$

From Theorem 3.8, for BVH$_2$, considering node (0,0) as the source node and (3,3) as destination node there are four node disjoint paths. Two of them include three processors with four links and the rest two with two processors with three links. So for BVH$_2$, using Eq. (7), we can the terminal reliability,

$$TR(BVH_2) = 1 - [(1 - R_l^m R_p^n)^k (1 - R_l^{m'} R_p^{n'})^{k'}] \qquad (8)$$

Now putting $R_l$=0.9; and $R_p$=0.8, Eq. (8) becomes

TR(BVH$_2$)= $1 - [(1 - 0.9^3 0.8^2)^2 (1 - 0.9^4 0.8^3)^2]$

=0.8745

### 5.4.3 Reliability of BVH$_3$

As stated earlier in Theorem 3.8, for BVH$_3$ considering node (000) as the source and (330) as destination there are six parallel paths. Four of them have four processors with five links and the rest two have two processors and three links. So the terminal reliability for BVH$_3$ is given by

TR(BVH$_3$)= $1 - [(1 - R_l^m R_p^n)^k (1 - R_l^{m'} R_p^{n'})^{k'}]$

$= 1 - [(1 - 0.9^5 0.8^4)^4 (1 - 0.9^3 0.8^2)^2]$

=0.9059

### 5.4.4 Reliability Analysis With Respect To Time

The reliability of a processor is calculated by

$$R_p(t) = e^{-\lambda_p t}, \qquad (9)$$

where $\lambda_p$ is the processor failure rate and $t$ is the mission time.

Similarly the link reliability is given by

$$R_l(t) = e^{-\lambda_l t}, \qquad (10)$$

where $\lambda_l$ is the link failure rate.

For the current work the link failure rate is assumed to be 0.0001 failures per hour and processor failure rate is assumed to be 0.001 failures per hour [4].

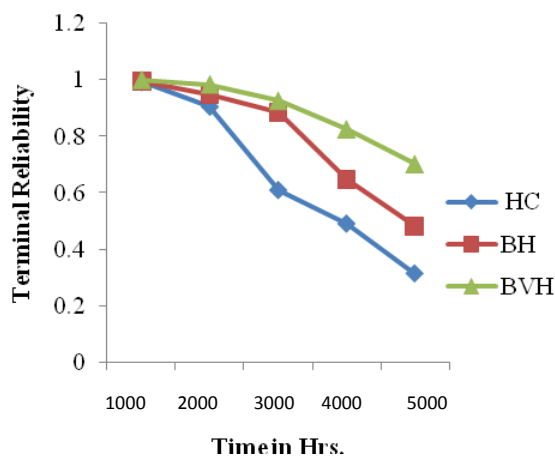

Figure 11: Comparison of Terminal Reliability for p=64

The Fig. 11 shows the comparative results of terminal reliability of Hypercube, Balanced Hypercube and Balanced varietal hypercube for a system having 64 numbers of processors. It is exponential in nature. It is clear from the Fig. 11 that the BVH is more reliable among all the three candidate networks.

## 6. CONCLUSION

This paper presented a new interconnection network topology called Balanced Varietal Hypercube for parallel systems. The new network is recursive and extensively hierarchical in structure. It retains most of the properties of both the balanced hypercube and varietal hypercube. Its properties are compared with that of hypercube, varietal hypercube, and balance hypercube. In terms of degree, diameter, cost, average distance and reliability, in general the proposed structure is shown to perform better than Hypercube, Varietal hypercube and Balanced hypercube.

**Authors**

Prof. (Dr.) C.R. Tripathy received the B.Sc. (Engg.) in Electrical Engineering from Sambalpur University and M. Tech. degree in Instrumentation Engineering from I.I.T., Kharagpur respectively. He got his Ph.D. in the field of Computer Science and Engineering from I.I.T., Kharagpur. He has more than 60 publications in different National and International Journals and Conferences. His research interest includes Dependability, Reliability and Fault–tolerance of Parallel and Distributed systems. He was recipient of "Sir Thomas Ward Gold Medal" for research in Parallel Processing. He is a fellow of Institution of Engineers, India. He has been listed as leading scientist of World 2010 by International Biographical Centre, Cambridge, England, Great Britain. He is also a Senior Member in Instrument Society of India, Orissa Information Technology Society and Life member in ISTE. He has three times received Best Paper award from Institution of Engineers, India.

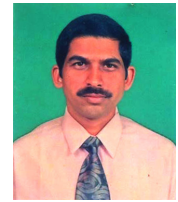

Ms. Nibedita Adhikari is working as Assistant Professor and Head, Department of Computer Science and Engineering, Purushottam Institute of Engineering and Technology, Rourkela, Orissa. She has received M. Tech degree in Computer Science and Engineering from National Institute of Technology Rourkela, MCA degree from G.M College (Auto) Sambalpur. She has also received Master of Sc. Degree in Mathematics form Utkal University Bhubaneswar. At present she is doing her PhD work at Sambalpur University. Her area of research is Parallel Computing, Performance Analysis and Interconnection Networks. She is an Associate Member of Institution of Engineers, India, Life Member of Computer Society of India and Orissa Information Technology Society.

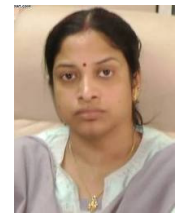